\documentclass{article}
\usepackage{spconf,amsmath,graphicx}
\usepackage[hidelinks]{hyperref}

\title{learning to fool the speaker recognition}
\name{Jiguo Li$^{1,2,4}$, Xinfeng Zhang${^{4}}$, Jizheng Xu$^3$, Li Zhang$^3$, Yue Wang$^3$, Siwei Ma$^{2\dagger}$, Wen Gao$^2$\thanks{The work was done when Jiguo Li interned in Bytedance.Inc.}\thanks{$^\dagger$Siwei Ma~(swma@pku.edu.com) is the corresponding author.}}
\address{$^1$Institute of Computer Technology, Chinese Academic of Sciences, \\$^2$Peking University, \\$^3$Bytedance.Inc, \\$^4$University of Chinese Academic of Sciences}
\begin{document}
%
\maketitle
\begin{abstract}
  Due to the widespread deployment of fingerprint/face/speaker recognition systems, attacking deep learning based biometric systems has drawn more and more attention.  Previous research mainly studied the attack to the vision-based system, such as fingerprint and face recognition. While the attack for speaker recognition has not been investigated yet, although it has been widely used in our daily life. In this paper, we attempt to fool the state-of-the-art speaker recognition model and present \textit{speaker recognition attacker}, a lightweight model to fool the deep speaker recognition model by adding imperceptible perturbations onto the raw speech waveform. We find that the speaker recognition system is also vulnerable to the attack, and we achieve a high success rate on the non-targeted attack. Besides, we also present an effective method to optimize the speaker recognition attacker to obtain a trade-off between the attack success rate with the perceptual quality. Experiments on the TIMIT dataset show that we can achieve a sentence error rate of $99.2\%$ with an average SNR $57.2\text{dB}$ and PESQ 4.2 with speed rather faster than real-time. 
\end{abstract}

\section{Introduction}
\label{sec:intro}
\begin{figure}[t]
  \centering
  \centerline{\includegraphics[width=.8\linewidth]{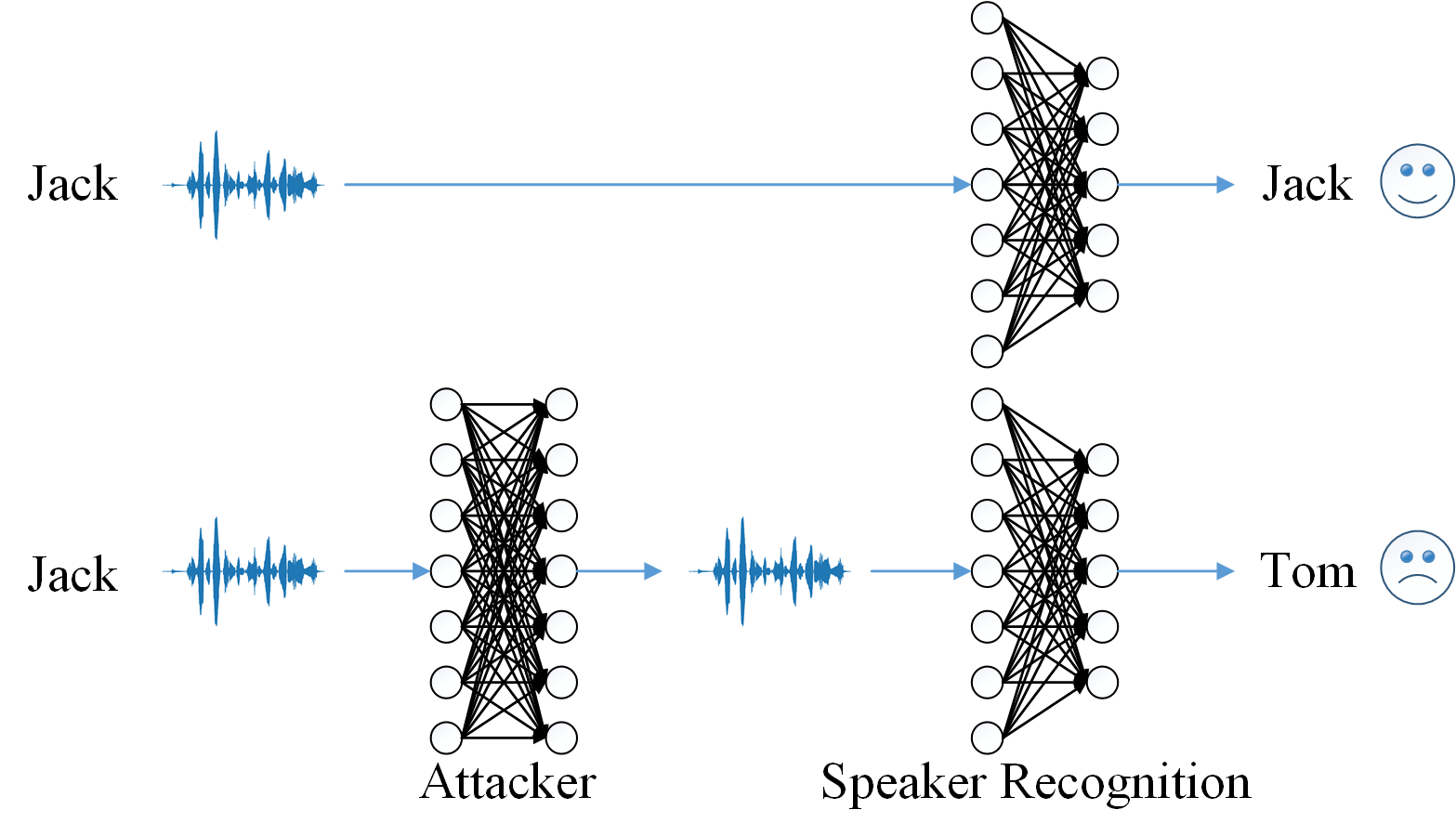}}
  \caption{Illustration of the Speaker Recognition Attacker}
  \label{fig:task}
\end{figure}

Deep network based biometric systems, such as fingerprint/face/speaker recognition, have been widely deployed in our daily life. Meanwhile, finding the weakness and attacking these recognition systems also draw more and more attention. Although many works have been done on vision-based systems~\cite{szegedy2013intriguing, goodfellow2014explaining,baluja2018learning}, 
the attack to speaker recognition has not been well-studied. There are two main applications of attacking the speaker recognition systems and finding the adversarial examples: (1) disturbing the speaker recognition systems when they are not wanted; (2) helping improve the performance and robustness of the speaker recognition systems. In this work, we focus on attacking the speaker recognition and present a model as well as its optimization method to attack the well-trained state-of-the-art deep speaker recognition model by adding the perturbations on the input speech, as illustrated in Fig.~\ref{fig:task}. 

\begin{figure*}[htp]
  \centering
  \centerline{\includegraphics[width=.75\linewidth]{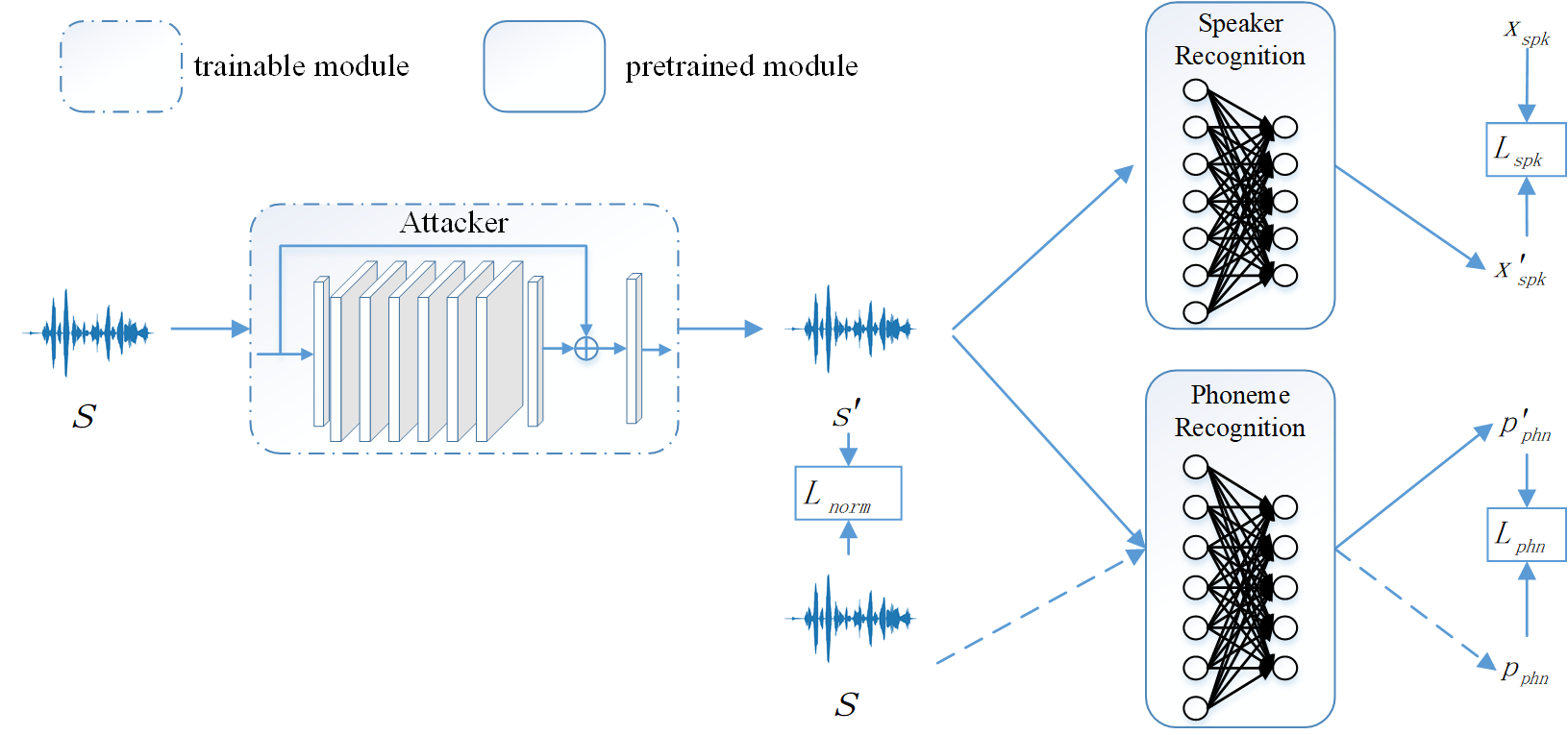}}
  \caption{Illustration of our framework. Our speaker recognition attacker is applied to the raw speech input and generates a speech with perturbations, which can fool the following speaker recognition model although the perturbations are imperceptible. Besides, a pretrained phoneme recognition model is used to help train the attacker network. No finetuning for the pretrained modules when training the attacker model.}
  \label{fig:model}
  \vspace{-3mm}
\end{figure*}

Attacking the deep neural networks~(DNNs) has become an emerging topic along with the development of DNNs since the weaknesses of DNNs have been found by Szegedy~\textit{et al.}~\cite{szegedy2013intriguing}. On the vision tasks, some optimization methods, such as L-BFGS~\cite{szegedy2013intriguing}, Adam~\cite{carlini2017towards}, or genetic algorithm~\cite{su2019one}, are used to modify the input pixels to obtain the adversarial examples. But these methods need the gradient or iterations during the testing phase, which are not practical in realistic scenarios. Baluja~\textit{et al.}~\cite{baluja2018learning} proposed adversarial transformation networks~(ATNs), which create a separated attacker network, to transform all inputs into adversarial ones. 

Base on the works on vision tasks, some methods are proposed to attack the automatic speech recognition~(ASR)  model. Alzantot~\textit{et al.}~\cite{alzantot2018did} proposed to attack the ASR model via a gradient-free genetic algorithm to generate adversarial examples iteratively. However, different from visual images, the psychoacoustic model shows that no difference will be perceived by humans if the distortion is under certain hearing thresholds. Therefore, Schonherr~\textit{at al.}~\cite{schonherr2018adversarial} and Szurley~\textit{et al.}~\cite{szurley2019perceptual} proposed to optimize the attack with the psychoacoustic model, and add perturbations under the hearing threshold. 
Our work is different from these audio attack works in two aspects. First, our work mainly focuses on the different task to attack the speaker recognition model. Second, our model is based on ATNs, which need no gradient during the testing phase and is fast for inference. Besides, although the replaying attack has been explored~\cite{korshunov2016overview}, learning based attack has not been well-studied.
Our contributions can be summarized as follows:
\begin{itemize}
  \item We attempt to attack the state-of-the-art speaker recognition model, and find that it is vulnerable to the attack.
  \item We propose a model to attack the speaker recognition model. In the non-targeted experiments conducted on TIMIT dataset~\cite{garofolo1993darpa}, we achieve a sentence error rate~(SER) of $99.2\%$ with the SNR up to $57.2$dB and PESQ up to 4.2 with speed rather faster than real-time.
  \item We present an optimization method to train our model, and experimental results show that our method can achieve a trade-off between the attack success rate and the perceptual quality.
\end{itemize}


\section{Speaker Recognition Attacker}
\label{sec:speaker_attacker}
As illustrated in Fig.~\ref{fig:model}, our proposed speaker recognition attacker, a trainable network, is used to add generative perturbations onto the input speech to attack the following pretrained speaker recognition model. A pretrained phoneme recognition model is used to help train the attacker. Given a speech $s$ and its speaker label $y_{spk}$, the non-targeted attack for the speaker recognition model can be formulated as:
\begin{align}
    \arg\min_{s'}{L(s, s')}&\quad\text{s.t.}\quad f(s')=y'_{spk}\\
    &\text{where}~s'=T_{f,\theta}(s)~\text{and}~y'_{spk}\neq y_{spk},\nonumber
\end{align}
where $T$ is the attack model to transform the input speech signal into an adversarial example, $f$ is a well-trained speaker recognition model, $L$ is a metric function to measure the distance between two samples~(e.g., the $L_2$ norm). The constraint $y'_{spk}\neq y_{spk}$ is changed to $y'_{spk}=y_{target}$ for the targeted attack.
\subsection{Attacker Network}
The proposed attacker network is a fully convolution residual network, including 5 convolution blocks totally in the residual branch, as illustrated in Fig.~\ref{fig:model}. 
1-D convolution, batchnorm, and ReLU are applied in every convolution block, following the setting in~\cite{baluja2018learning}. The kernel size for all convolution layers is set as $3$ and the channel is set to be $32$. To increase the receptive field, we use different dilations in the different convolution layers. The dilations for 5 convolution layers are $1,2,5,2,1$, respectively. Besides, we init the weight and bias of the last convolution layer as zero so that our model adds no perturbation at the start of the training, which is important for the optimization to keep the perturbations on a small scale.
\subsection{Optimization}
The intuitive method to train the attacker network is gradient ascent, however, in practice, it fails because a well-trained speaker recognition model propagates back almost zero gradient due to the softmax layer. Motivated by the Wasserstein GAN~\cite{arjovsky2017wasserstein} to optimize the Wasserstein distance between two distributions, we just solve this gradient missing problem to optimize directly on the immediate activation before the softmax layer. On the other hand, we also need to ensure the perturbations are imperceptible. $L_2$ norm is used to constraint the scale of the perturbations. Besides, we also take the phoneme information into the account via a pretrained phoneme recognition network to optimize the perceptual quality. In summary, we optimize our attacker network from three aspects:
\begin{align}
  L_{total}&=L_{spk}+\lambda_{phn}L_{phn}+\lambda_{norm}L_{norm}\\
  L_{spk}&=\left\{
    \begin{aligned}
    x'_{spk}[I_{1st}]-x'_{spk}[I_{2nd}]&,\quad I_{1st}=y_{spk} \\
    0&,\quad \text{else}\\
    \end{aligned}
  \right.\\
  L_{phn}&=\text{KL}(p_{phn}\|p'_{phn})\\
  L_{norm}&=[\max(s-s'-m, 0)]^2,
\end{align} 
where $x'$ is the immediate activation before the softmax layer of the speaker recognition model with the input $s'$, $p/p'$ is the output distribution of the softmax layer of the phoneme recognition model with the input $s/s'$, $I_{1st}/I_{2nd}$ is the index of 1st/2nd largest value in $x'_{spk}$. In $L_{phn}$, we use Kullback–Leibler divergence~(KLD) to measure the distance between two distributions. In $L_{norm}$, $m$ is a hyper-parameter to give a margin in which the perturbations are thought imperceptible. $\lambda_{phn}$ and $\lambda_{norm}$ are used to fuse the three loss items. For the targeted attack, the loss is the same except that $L_{spk}$ is changed to
\begin{align}
  L_{spk,target}&=\left\{
    \begin{aligned}
    x'_{spk}[I_{1st}]-x'_{spk}[y_{target}]&,\quad I_{1st}\neq y_{target} \\
    0&,\quad \text{else,}\\
    \end{aligned}
  \right.
\end{align} 
where $y_{target}$ denotes the target speaker class.

\subsection{Inference}
The input speeches with variable length are split into frames with fixed length due to the fully connection layer in the speaker recognition model in the training stage. However, in the testing stage, the input speeches can be of arbitrary length because our attacker network is a fully convolution residual network. The inference is fast because our attacker network is lightweight with only 5 convolution blocks and small kernel size filters.

\begin{table}
  \begin{center}
  \begin{tabular}{c|c|c|c|c}
    \hline
    $\lambda_{phn}$ & $\lambda_{norm}$ & SER($\%$)$\uparrow$ & SNR(dB)$\uparrow$ & PESQ$\uparrow$ \\
    \hline 
    \hline
     - & - & $1.52^\star$ &  - & -  \\
     \hline
     0 & 0 & 99.7 & 18.56 & 1.09 \\
     0 & 1000 & 96.5 & 56.39 & 3.72 \\
     0 & 2000 & 86.7 & 57.79 & 3.61 \\
     \hline
     1 & 1000 & \textbf{99.2} & 57.20 & 4.20  \\
     5 & 1000 & 93.9 & 58.00 & 4.25 \\
     10 & 1000 & 90.5 & \textbf{59.01} & \textbf{4.28} \\
    \hline
  \end{tabular}
  \caption{Non-targeted attack results. The first row is the baseline performance of speaker recognition without attack. The rest of the table shows the attack results of our model with different trade-offs between SER with SNR and PESQ by tuning $\lambda_{phn}$ and $\lambda_{norm}$. $^\star$This result is not the same as that in~\cite{ravanelli2018speaker}, but this model is released by the author.}
  \label{tab:non_target_attack}
  \end{center}
  \end{table}

\section{Experimental Result}
\label{experiment}

\subsection{Experimental Setup}
\textbf{Pretrained Speaker/Phoneme Recognition Model.} We use the state-of-the-art speaker recognition model, SincNet~\cite{ravanelli2018speaker}, as the target model to attack. SincNet replaces the first layer of a CNN with a group of learnable bandpass filters. In this way, the network is more interpretable and show better performance~\cite{ravanelli2018interpretable}. Besides, SincNet also works on phoneme recognition\footnote{https://github.com/mravanelli/pytorch-kaldi}. In our experiment, we use the official released pretrained SincNet model for speaker recognition. The phoneme recognition model is referred from Pytorch-Kaldi~\cite{pytorch-kaldi} and achieve a $26.4\%$ frame error rate on the TIMIT\cite{garofolo1993darpa} dataset\footnote{we use the same train/test split for phoneme and speaker recognition, which is different with the typical split manner for phoneme recognition.}. \\
\textbf{Dataset and Metric.} Following the setting in~\cite{ravanelli2018speaker}, we conduct experiments on TIMIT~(462 speakers, train chunk)~\cite{garofolo1993darpa} to demonstrate the performance of our proposed model.  The signal-to-noise ratio~(SNR) and Perceptual Evaluation of Speech Quality~(PESQ) score~\cite{rix2001perceptual} are used to evaluate the objective and perceptual quality, respectively. SNR is calculated as follows:
\begin{align}
  \text{SNR}=10\log_{10}[(\sigma_s^2/\sigma_e^2)]^2,
\end{align}
where $\sigma_s^2/\sigma_e^2$ is the mean square of input signal/error. PESQ~\cite{rix2001perceptual} is an integrated perceptual model to map the distortion to a prediction of subjective mean opinion score (MOS) with range $[0.5,4.5]$, which is an ITU-T recommendation technology~\cite{itu2001pesq}. Following the works for attacking image classification~\cite{szegedy2013intriguing,goodfellow2014explaining}, we use the classification error rate~(CER) to evaluate the performance of our attacker. For the non-targeted attack, sentence error rate (SER), which is defined as the CER of the speech sentences, is used to measure our attacker's performance. For the targeted attack, our attack is successful as long as the prediction is the target, so we use prediction target rate (PTR), which is the percentage of the target in the predictions over the testing set, to measure our attacker's performance.
\\
\textbf{Trainging Details.}  The speech sentences, with sampling rate 16k, are split into 200ms frames with 10ms stride, following~\cite{ravanelli2018speaker}. The data will be normalized before being feed into the attacker model and de-normalized when they output the attacker model. After finetuning the hyper-parameters, we set $\lambda_{phn}=1$, $\lambda_{norm}=1000$ and $m=0.01$. We use Adam~\cite{kingma2014adam} optimizer with a learning rate of $3\times 10^{-4}$ to train the attack model for $10$ epochs. Data, code, and pretrained models have been released on our \href{https://smallflyingpig.github.io/speaker-recognition-attacker/main}{project home page}\footnote{https://smallflyingpig.github.io/speaker-recognition-attacker/main}.
\subsection{Non-targeted Attack}

To demonstrate the effectiveness of our proposed model, we conduct non-targeted attack experiments on the TIMIT dataset. The results are illustrated in Table.~\ref{tab:non_target_attack}, some conclusions can be drawn:
\begin{itemize} 
  \item Our proposed model successfully attacks the trained state-of-the-art speaker recognition model with an SER of $99.2\%$ on the testing set. Meanwhile, the perturbations are small enough to be imperceptible for humans because the SNR is up to 57.2dB and PESQ is no less than 4.2, indicating the efficiency of our model. 
  \item Tuning $\lambda_{norm}$ fails to get a trade-off between SER and PESQ~(row 1st, 2nd, 3rd in Table.~\ref{tab:non_target_attack}), while tuning $\lambda_{phn}$ works for that~(2nd, 4th$\sim$6th in Table.~\ref{tab:non_target_attack}), demonstrating our optimization is effective to achieve a trade-off between the attack success rate and and the perceptual quality.
  \item The results in the 2nd row and 4th row show that $L_{phn}$ can improve the performance of the attacker model~(measured by SER), as well as the quality of the adversarial examples~(measured by SNR and PESQ).
\end{itemize}


\begin{figure}[t]
  \centering
  \begin{minipage}[b]{0.85\linewidth}
    \centering
    \centerline{\includegraphics[width=.99\linewidth]{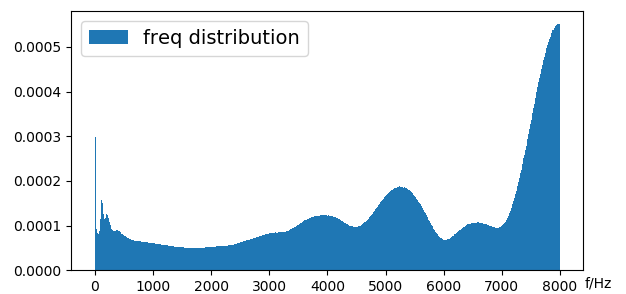}}
  \end{minipage}
  \caption{
    The spectrogram energy distribution of the perturbations over the testing set.}
  \label{fig:dist}
  
  \end{figure}


Besides, we also give the distribution of the perturbations on the frequency domain to study if the perturbations have frequency selectivity.
The frequency distributions of the perturbations from the attacker model with $\lambda_{phn}=1$ are shown in Fig.~\ref{fig:dist}. The spectrogram energy distribution shows that: (1) the perturbations are full-band, and all the frequencies are useful for the attack; (2) the energy in the 7k$\sim$8k band is significantly stronger than that in other bands, indicating the high-frequency band has much affect on the speaker recognition's performance. The frequency characteristics of perturbations have a great influence on the model performance and signal quality, but looking deeper into this question is beyond the scope of this paper.
\subsection{Targeted Attack}
Besides the non-targeted attack, we also evaluate our model on the targeted attack. We fix the hyper-parameters as $\lambda_{norm}=1000, \lambda_{phn}=1, m=0.01$ in the targeted attack. Five speakers are randomly selected from the 462 speakers of the TIMIT dataset as targets. The attack results are illustrated in Table.~\ref{tab:target_attack}. The results show that:
\begin{itemize}
  \item Our model can attack the speaker recognition with a PTR of $72.1\%$ on average over the five targets. Meanwhile, the perturbations are small enough because the SNR is up to $57.64$dB.
  \item The PESQ (3.48 on average) is well enough although it is not as good as that in the non-targeted attack, given the fact that targeted attack is more challenging than the non-targeted attack.
\end{itemize}

\begin{table}
    \begin{center}
    \begin{tabular}{c|c|c|c}
      \hline
      Target ID  & PTR($\%$)$\uparrow$ & SNR(dB)$\uparrow$ & PESQ$\uparrow$ \\
      \hline 
      \hline
       0 & 91.4 & 57.55 & 3.36  \\
       100 & 89.3 & 56.83 & 3.16 \\
       200 & 63.3 & 58.42 & 3.69   \\
       300 & 58.7 & 56.92 & 3.52  \\
       400 & 57.6 & 58.36 & 3.68 \\
      \hline
       avg & 72.1 & 57.64 & 3.48 \\
      \hline
    \end{tabular}
    \caption{Targeted attack results. The first five rows are the results of the targeted attack for five randomly selected speakers from the TIMIT dataset. The last row is the average result of the above five rows. PTR denotes the prediction target rate.}
    \label{tab:target_attack}
    \end{center}
    \end{table}
\subsection{Real-time Attack}
Our attack model is very lightweight with only 5 convolution blocks, so it is very fast. To verify it is fast enough to process the speech data in real-time, we calculate the real-time factor(RTF) over the testing set. RTF is defined as the ratio of the processing time to the input duration and the system is real-time if $\text{RTF}\leq 1$.
We test our model in CPU mode on a machine with an Inter(R) Core(TM) i7-6700K @ 3.4GHz CPU and get an average RTF 0.042, indicating that our attacker is more than 20 times faster than the real-time requirement, although it runs in the CPU mode. 
\section{conclusion}
\label{conclusion and future works}
In this paper, we proposed a model to attack the speaker recognition model by training a lightweight attacker network to add perturbations on the input speech. Experiments show that our model was effective and efficient on both non-targeted and targeted attacks. We have built a pioneer work on the learning based speaker recognition attack and established the corresponding benchmark for such study. In the future, the black-box attack and transferable attack will be explored because the gradient of the target is usually inaccessible in the real scenarios.

\vfill\pagebreak

\bibliographystyle{IEEEbib}
\bibliography{strings,refs}

\end{document}